\documentclass[12pt]{emulateapj}
\slugcomment{{\it Accepted for publication in ApJL.}}

\usepackage{epsfig}
\usepackage{amsmath}
\usepackage{amssymb}
\usepackage{natbib}
\usepackage{graphicx}

\def\Mpc{{\rm\thinspace Mpc\ }}   
\def\kms{{\rm\thinspace km\thinspace s}^{-1}}
\def\cm{{\rm\thinspace cm}}

\def\kpc{{\rm\thinspace kpc}}

\def\Mpc{{\rm\thinspace Mpc\ }}   
\def\Msun{\hbox{$\mathrm{\thinspace M_{\odot}}$}}

\def\yr{{\rm\thinspace yr}}     
\def\Myr{{\rm\thinspace Myr}}     
\def\Gyr{{\rm\thinspace Gyr}}     
\def\HI{{\rm H\textsc{i}\ }}

\def\Gadget2{\rm{\textsc{Gadget\thinspace 2}\ }}

\shorttitle{Modeling Star Formation in the Antennae}
\shortauthors{Karl et al.}

\begin{document}

\title{One moment in time - modeling star formation in the Antennae}
\author{Simon J. Karl$^1$, Thorsten Naab$^1$$^,$$^2$, Peter H. Johansson$^1$,
  Hanna Kotarba$^1$, Christian M. Boily$^3$, Florent Renaud$^3$$^,$$^4$, Christian Theis$^4$$^,$$^5$}
\affil{$^1$ Universit\"ats-Sternwarte M\"unchen, Scheinerstr.\ 1, D-81679 M\"unchen,  
Germany;\\
\texttt{skarl@usm.lmu.de}}
\affil{$^2$ Max-Planck-Institut f\"ur Astrophysik,
  Karl-Schwarzschild-Str. 1, D-85741 Garching bei M\"unchen,
  Germany;\\ \texttt{naab@mpa-garching.mpg.de}}
\affil{$^3$ Observatoire astronomique, Universit\'e de Strasbourg and CNRS UMR 7550,\\
11 rue de l'Universit\'e, F-67000 Strasbourg, France}
\affil{$^4$Institut f\"ur Astronomie der Univ. Wien, T\"urkenschanzstr. 17, A-1180 Vienna, Austria}
\affil{$^5$ Planetarium Mannheim, Wilhelm-Varnholt-Allee 1, D-68165
  Mannheim, Germany}

\begin{abstract}
We present a new high-resolution N-body/SPH simulation of an encounter of two 
gas-rich disk galaxies which closely matches the morphology and kinematics of the interacting 
Antennae galaxies (NGC 4038/39). The simulation includes radiative cooling, star formation and feedback
from SNII. The large-scale morphology and kinematics are determined by the internal structure 
and the orbit of the progenitor disks. The properties of the central region, in particular the starburst
in the overlap region, only match the observations for a very short time interval ($\Delta t \approx$ 20
\Myr) after the second encounter. This indicates that the Antennae galaxies are in a special phase only
about 40 $\Myr$ after the second encounter and 50 $\Myr$ before their final collision. 
This is the only phase in the simulation when a gas-rich overlap region between the nuclei is 
forming accompanied by enhanced star formation. 
The star formation rate as well as the recent star formation history in 
the central region agree well with observational estimates. For the first time this new model explains
the distributed extra-nuclear star formation in the Antennae galaxies
as a consequence of the recent 
second encounter. The proposed model predicts that the Antennae are in a later merger stage than the Mice
(NGC 4676) and would therefore lose their first place in the classical Toomre sequence.  
\end{abstract}


\section{Introduction}
\begin{figure*}
\centering 
\plotone{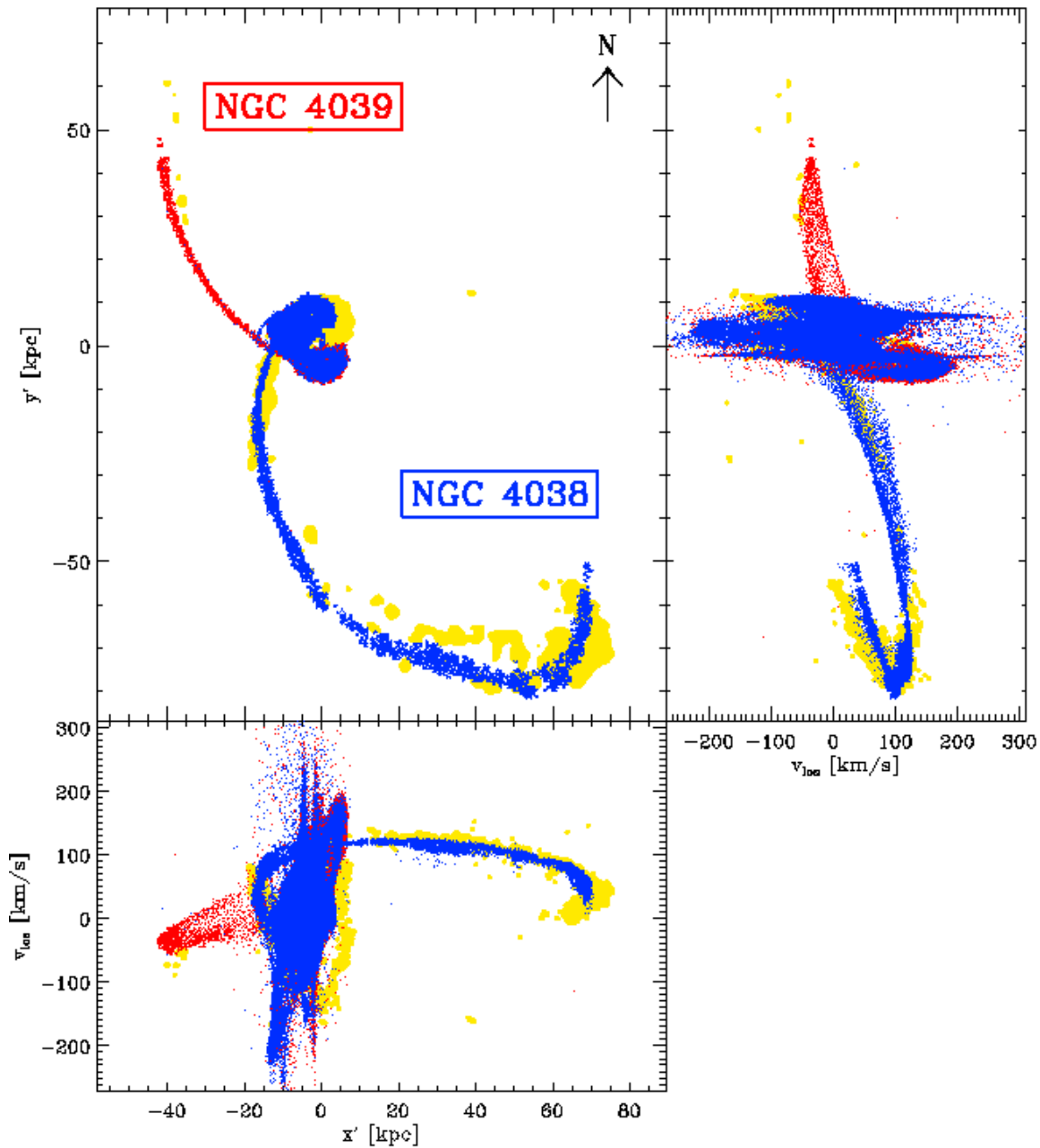}
\caption{Simulated gas properties projected on top of  \HI
  kinematic data by \citet{HibbardEtAl2001AJ} at the time of best
  match. Simulated gas particles are displayed in blue (NGC 4038) and red
  (NGC 4039). Yellow points represent the observational data. 
  {\it Upper left panel:} Projected positions in the
  plane-of-the-sky (x$^\prime$-y$^\prime$ plane). {\it Upper right and lower left panel:} 
  Declination (y$^\prime$) and Right Ascension (x$^\prime$) against line-of-sight velocity. 
  Similarly to the observations, we apply a column density threshold
  of $N_\mathrm{gas} = 10^{20} \cm^{-2}$ to the simulated gas distribution.}
\label{pic1:PV}
\end{figure*}
\label{Intro}
In the local universe ($z < 0.3$) about $\sim5-10\%$ of all galaxies are interacting 
and merging (e.g. \citealp{2008ApJ...672..177L,2010ApJ...709.1067B}).
Mass assembly via this mechanism was more important at earlier cosmic times
when major mergers were more frequent
\citep[e.g.][]{2002ApJ...565..208P,ConseliceEtAl2003AJ....126.1183C} and also
more gas-rich \citep[e.g.][]{2010Natur.463..781T}. Major mergers dramatically affect the formation and evolution 
of galaxies. By inducing tidal torques they can efficiently
transport gas to the centers of the galaxies
\citep{BarnesHernquist1996ApJ, 2006MNRAS.372..839N}, trigger star
formation \citep{Mihos&Hernquist1996ApJ, 2000MNRAS.312..859S,
  2008MNRAS.384..386C}, feed super-massive black holes
\citep{2005ApJ...630..705H, SpringelDiMatteoHernquist2005MNRAS, JohanssonEtAl2009ApJ}
and convert spiral galaxies into intermediate-mass ellipticals 
\citep{1992ApJ...393..484B, NaabBurkert2003ApJ,2004AJ....128.2098R,2009ApJ...690.1452N}.

The \object{Antennae} galaxies (NGC 4038/39) are the nearest and
best-studied example of an on-going major merger of two gas-rich
spiral galaxies. The system sports a beautiful pair of elongated tidal tails
extending to a projected size of $\sim20 \arcmin$ (i.e. $106 \kpc$ at
an assumed distance of 22 Mpc), 
together with two clearly visible, still distinct galactic disks. 
The latter has been assumed to be an indication of an early merger state, putting the
system in the first place of the \citet{Toomre1977egsp.conf..401T}
merger sequence of 11
prototypical mergers. Due to their proximity and the ample number of high-quality
observations covering the spectrum from radio to
X-ray 
\citep[e.g.][]{NeffUlvestad2000AJ, WangEtAl2004ApJS, WhitmoreEtAl1999AJ,
  2005ApJ...619L..87H, ZezasEtAl2006ApJS..166..211Z}
the Antennae provide an ideal laboratory for understanding the
physics of 
merger-induced starbursts through comparison with high-resolution simulations. 

At the center of the Antennae galaxies, HST imaging has revealed a large
number of bright young star clusters ($\gtrsim 1000$) which plausibly 
have formed in several bursts of star formation induced by the
interaction \citep{WhitmoreEtAl1999AJ}.
The spatial distribution and the age of these clusters are correlated:
the youngest clusters are found in the overlap
region ($\tau < 5 \Myr$), while the young starburst is generally
located in the overlap and a ring-like configuration in the disk of
NGC 4038 ($\tau \lesssim 30 \Myr$). An intermediate-age population
($\tau = 500-600 \Myr$) is distributed throughout the disk of NGC 4038
\citep{WhitmoreEtAl1999AJ,ZhangFallWhitmore2001ApJ}. 

Of particular interest is the spectacular nature of an extra-nuclear starburst
observed in the dusty overlap region between the merging galactic 
disks \citep{MirabelEtAl1998A&A,WangEtAl2004ApJS}. The Antennae seem to
be the only interacting system where an off-center starburst is 
outshining the galactic nuclei in the mid-IR
\citep{XuEtAl2000ApJ} and among only a few systems which show
enhanced inter-nuclear gas concentrations \citep{1999ApJ...524..732T}. To
date, this prominent feature has not been reproduced in any simulation
of the Antennae system \citep[see][]{BarnesHibbard2009AJ}. Thus, the question 
remains whether this feature cannot be captured by current sub-grid
modeling of star formation or whether the previous dynamical models
(e.g. initial conditions) were not accurate enough.

A first simulation of the Antennae galaxies was presented by 
\citet{Toomre&Toomre1972ApJ}, reproducing the correct trends in the
morphology of the tidal tails.
\citet{Barnes1988ApJ} repeated the analysis with a self-consistent
multi-component model consisting of a bulge, disk and dark halo component. 
\citet{MihosBothunRichstone1993ApJ} included gas and star formation 
in their model and found the star formation to be concentrated 
in the nuclei of the disks, thus, not reproducing the overlap star formation.

In this Letter, we present the first high-resolution merger
simulation of NGC 4038/39 with cosmologically motivated 
progenitor disks galaxy models. We are able to match both the large-scale
morphology and the line-of-sight kinematics, as well as
important key aspects of the distribution and ages of newly-formed
stars at the center of the Antennae, being a direct consequence of the
improved merger orbit.

\section{Simulations} 
\label{modelling}
The simulation presented here is the best-fitting model of a larger
parameter study and was performed using 
\Gadget2 \citep{Springel2005MNRAS}. We include primordial radiative cooling 
and a local extra-galactic UV background. 
Star formation and associated SNII feedback are modeled following the sub-grid multi-phase prescription
of \citet{Springel&Hernquist2003MNRAS}, but excluding
supernovae-driven galactic winds. For densities $n > n_\mathrm{crit} =  0.128 \cm^{-3}$
the ISM is treated as a two-phase medium with cold clouds embedded in
pressure equilibrium in a hot ambient 
medium.  We deploy a fiducial set of parameters 
governing the multi-phase feedback model resulting in a star 
formation rate (SFR) of $\sim1 \Msun \yr^{-1}$ for a Milky Way-type galaxy.
We adopt a softened equation of state (EQS) with $q_\mathrm{EQS} =
0.5$, where the parameter $q_\mathrm{EQS}$ interpolates the star formation model
between the full feedback model ($q_\mathrm{EQS}=1.0$) and an isothermal EQS with 
$T=10^{4} \ \rm K$ ($q_\mathrm{EQS}=0$) (see \citealp{SpringelDiMatteoHernquist2005MNRAS} 
for further details).
\begin{table}
\caption{Model parameters of the best-fit merger configuration}
\label{Tab:RunParameters}      
\centering          
\begin{tabular}{ c | c | c }
\hline
\hline                             
Property & \object{NGC 4038} & \object{NGC 4039} \\
\hline                             
$M_\mathrm{vir}$\footnote{Mass in
  $10^{10}\Msun$} & $55.2$ & $55.2$ \\
$M_{\mathrm{disk, stellar}}$ & $3.3$ & $3.3$ \\
$M_{\mathrm{disk, gas}}$ & $0.8$ & $0.8$ \\
$M_{\mathrm{bulge}}$ & $1.4$ & $1.4$ \\
$r_{\mathrm{disk}}$\footnote{Disk and bulge lengths
  ($r_{\mathrm{disk}}$, $r_{\mathrm{bulge}}$) and disk scale height ($z_0$) are given in
  $\kpc$} & $6.28$ & $4.12$ \\
$z_0$ & $1.26$ & $0.82$ \\
$r_{\mathrm{bulge}}$ & $1.26$ & $0.82$ \\
c\footnote{Halo
  concentration parameter} & $15$ & $15$ \\
$\lambda$\footnote{Halo
  spin parameter} & $0.10$ & $0.07$ \\
$v_{\mathrm{rot}}^{\mathrm{max}}$\footnote{Maximum rotational velocity in $\kms$} & $189$ & $198$\\
\hline
\hline
\end{tabular}
\end{table}
The progenitor galaxies are set up in equilibrium according to the method of
\citet{SpringelDiMatteoHernquist2005MNRAS} with a total virial mass of 
$M_{\mathrm{vir}} = 5.52 \times 10^{11} \Msun$ for each galaxy. The dark matter halos are 
constructed using a \citet{Hernquist1990ApJ} density profile.  They are 
populated with exponential stellar disks comprising a constant disk mass fraction
$m_{\mathrm{d}} = 0.075$ of the total virial
mass and a stellar Hernquist bulge with a bulge mass fraction of 
$m_{\mathrm{b}} = 0.025$ $(m_{\mathrm{b}}=1/3 m_{\mathrm{d}})$. The gas mass fraction of the 
disk component is $f_{\mathrm{g}} = 0.2$ with the rest of the disk mass remaining in stars.
The disk and bulge scale lengths are determined using the
\citet*{MoMaoWhite1998MNRAS} formalism. A summary of the most relevant model 
parameters is given in Table \ref{Tab:RunParameters}. 

Each galaxy is realized with $N_\mathrm{tot} = 1.2 \times 10^6$ particles, 
i.e. 400,000 halo particles, 200,000 bulge particles, 
480,000 disk particles and 120,000 SPH particles. In order to avoid spurious
two-body effects we ensured that all baryonic 
particles have the same mass and only {\it one} stellar particle is spawned per SPH particle.
This yields a total baryon fraction of $f_\mathrm{bary} = 10\%$ with particle masses for the baryonic
components (bulge, disk, formed stars, and gas) of $m_\mathrm{bary} = 6.9 \times 10^4 \Msun$ and $m_\mathrm{DM} = 1.2
\times 10^6 \Msun$ for the dark matter particles.

The gravitational softening lengths are set to $\epsilon_\mathrm{bary} = 0.035 \kpc$ for baryons, 
and $\epsilon_\mathrm{DM} = 0.15 \kpc$ for dark matter particles, scaled according to 
$\epsilon_\mathrm{DM} = \epsilon_\mathrm{bary} \thinspace(m_\mathrm{DM}/m_\mathrm{bary})^{1/2}$.

We adopt an initially nearly-parabolic, prograde orbit geometry (the orbital 
plane lies in the x-y plane)
with a pericenter distance of $r_{\mathrm{p}} = r_{\mathrm{d,4038}} + r_{\mathrm{d,4039}} = 10.4 \kpc$ 
and an initial separation of $r_{\mathrm{sep}} = r_{\mathrm{vir}} = 168 \kpc$. For the orientation 
of the progenitor disks we found the best match to the Antennae system with 
inclinations $i_{\mathrm{4038}} = 60\degr,\, i_{\mathrm{4039}} = 60\degr$ and arguments of pericenter
$\omega_{\mathrm{4038}}= 30\degr,\, \omega_{\mathrm{4039}}= 60\degr$
\citep[see][]{Toomre&Toomre1972ApJ}.

\section{Results}
\label{results}
\subsection{The morphological and kinematical match}
\label{subsec:KinModel}

We determine the time when the simulation best matches the Antennae 
together with the viewing angles ($\theta$,$\psi$,$\phi$) which
specify a series of subsequent rotations around the x-, y-, and
z-axis. In the further analysis we will use the rotated 3D 
position-velocity subspace, i.e. the
plane-of-the-sky (x$^\prime$-y$^\prime$ plane) and the line-of-sight velocity
$v_{\mathrm{los}}$, for comparison with the observations. Finally, we apply a distance scale ($\mathcal{L}$) relative
to a fiducial distance of 22 \Mpc \citep{SchweizerEtAl2008AJ}\footnote{Note that the distance to 
  the Antennae is a matter of recent debate. The systemic recession
  velocity yields a distance of $19.2 \Mpc$ (assuming
  $\mathrm{H}_0=75\,\kms\,\Mpc^{-1}$) while photometry of the tip of the
  red giant branch suggests a much shorter distance of only $13.3
  \Mpc$ \citep{SavianeEtAl2008ApJ...678..179S}. Recently,
  \citet{SchweizerEtAl2008AJ} have used three independent methods to determine a distance of $22\pm 3 \Mpc$.} and
assume a systemic helio-centric velocity of $1630 \kms$ to fit
the observational data to the physical scales in the simulation.

We find our best match to the observed large- and small-scale
properties of the system with viewing angles of $(93,69,253.5)$ and
$\mathcal{L} = 1.4$, yielding a distance of $D = 30.8 \Mpc$ to the
system. The "best fit" ($t=1.24 \Gyr$ after beginning
of the simulation) is reached only $\sim40 \Myr$ after
the second encounter ($t = 1.20 \Gyr$), and approximately $50 \Myr$
before the final merging of the galaxy centers ($t = 1.29
\Gyr$). From our larger parameter study we found this exact
timing to be a mandatory requirement for reproducing the overlap starburst.
The first close passage of
the two progenitor disk galaxies occurred $\sim600 \Myr$ ago which is
considerably longer ago than $\sim200\, -\, 400 \Myr$ as suggested in
earlier models \citep{Barnes1988ApJ, MihosBothunRichstone1993ApJ} and
in much better agreement with observed 'intermediate-age' star clusters
($\sim500-600 \Myr$).
\begin{figure*}
\centering 
\includegraphics[width=18cm]{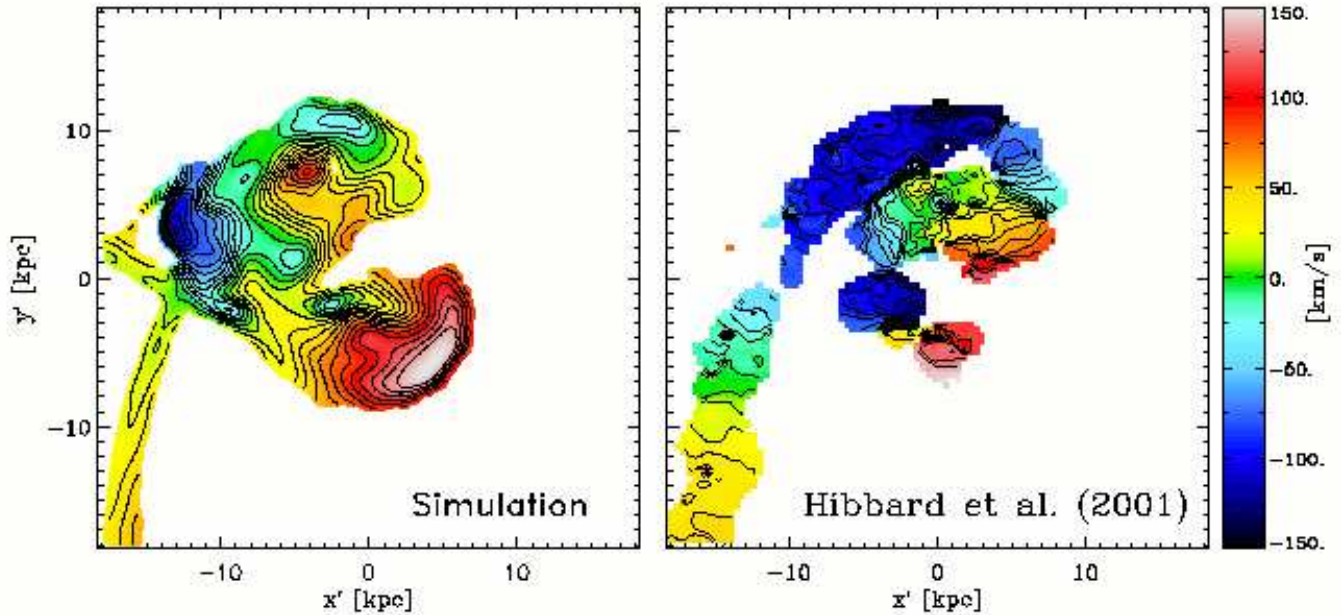}
\caption{Line-of-sight velocity fields inside $18 \kpc$ of the simulated and 
  observed central disks. Isovelocity contours are drawn at $10 \kms$
  intervals ranging from $-150 \kms$ to $150 \kms$. {\it Left:}
  density-weighted velocity map. {\it Right:} intensity-weighted 
  \HI velocity field of the high-resolution
  data cube. A column density threshold is applied as in
  Fig.\ref{pic1:PV}.}
\label{pic2:LoSHIVelfield}
\end{figure*}
In Fig.\ref{pic1:PV} we show three large-scale projections of the PV
cube of our simulated gas particles (NGC 4038: blue, NGC 4039: red) together with a 
direct comparison to \HI observations (yellow) by \citet{HibbardEtAl2001AJ}. 
The \HI gas phase is used here as a tracer for the smooth underlying morphological and
kinematical structure of the gas in the Antennae and we apply,
similarly to the \HI observations a column density
threshold of $N_{\mathrm{\HI}} \leq 1\times 10^{20}\cm^{-2}$ in the simulation. 
The top left panel displays the plane-of-the-sky projection, while in the
top right and bottom left panels we show two orthogonal
position-velocity profiles, Declination versus $v_\mathrm{los}$ (upper
right) and $v_\mathrm{los}$ versus Right Ascension (lower left). The
simulation matches the morphology and kinematics of the observed
system very closely, especially for the southern arm, including the
prominent kink in the velocities at the tip of the tidal arm (see
Fig.\ref{pic1:PV}, upper right and lower left panels). Due to the different
initial orientations of the progenitor disks, the gas distribution in
the northern arm is more diffuse than in the southern arm. The assumed
column density cut-off therefore results in a similar characteristic
stubby geometry as observed (Fig.\ref{pic1:PV}, upper left).

A closeup of the simulated and observed line-of-sight gas velocity fields in the
central $18 \kpc$ of NGC 4038/39 is shown in the left and right panels of 
Fig.\ref{pic2:LoSHIVelfield}. Gas particles are binned on a
  SPH-kernel weighted $256^3$ grid and summed up along the
  line-of-sight to produce a density-weighted velocity map
  \citep[see][]{HibbardEtAl2001AJ}. The grid is smoothed with the
  observed beam profile and displayed using the same projected pixel
  sizes ($\Delta_\mathrm{RA} = 2.64 \arcsec$ and $\Delta_\mathrm{Decl}
  = 2.5 \arcsec$) as in \citet{HibbardEtAl2001AJ}. We overlay
  isovelocity contours spaced by $10 \kms$ and apply the same column
  density threshold as in Fig.\ref{pic1:PV}. The simulation agrees
well with the observed velocity field of the disk of NGC 4038. The northern part is 
approaching and the southern part is receding at similar velocities. 
Similarly, the simulated disk of NGC 4039 is approaching in the northern part 
and receding in the southern part like in the observations. In the simulation we have 
significantly more gas in the overlap region and the southern disk
than in the observed $\HI$ velocity field. This is due to the fact that we do not distinguish between 
molecular and atomic gas in our simulation whereas most of 
the gas in the central regions of the Antennae seems to be in molecular
form \citep{GaoEtAl2001ApJ}.

\subsection{The recent starburst}
\begin{figure}
\centering 
\includegraphics[width=9cm]{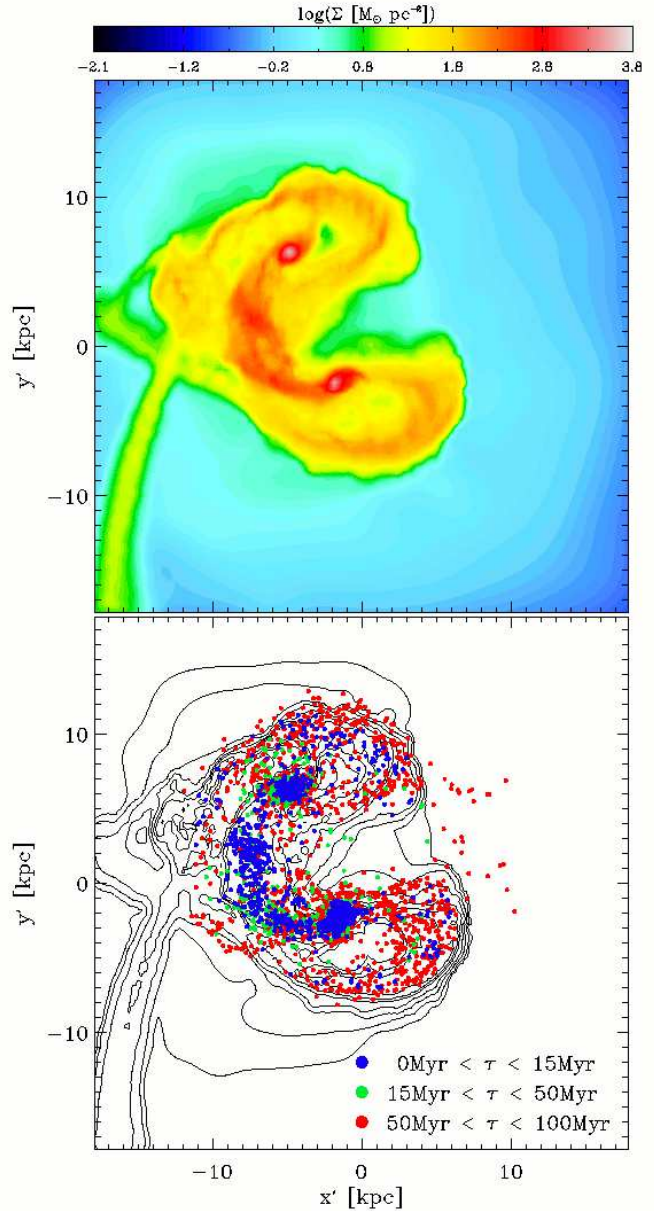}
\caption{{\it Top:} Gas surface density in the central $18 \kpc$ of the simulation. There are clear 
concentrations of gas in at the two centers of the galaxies and the overlap region (red contours).
{\it Bottom:} Recently formed stellar particles color-coded by their ages. The youngest stars 
(blue: $\tau < 15 \Myr$) have formed predominantly in the
  overlap region and the centers associated with the peaks in the gas surface density, and
  tidal features around the disks (see upper panel). 
  Older stars (green: $15 \Myr < \tau < 50 \Myr$; red: $50 \Myr < \tau <
  100 \Myr$) have formed throughout the galactic disks and the tidal arcs. 
}
\label{pic3:RemnantDisks}
\end{figure}

In Fig.\ref{pic3:RemnantDisks} we show a color-coded map of the total gas surface density
in the central $18 \kpc$ of the simulation (upper panel). The nuclei of the progenitor disks 
are still distinct and connected by a bridge of high density gas. In the lower panel 
of Fig.\ref{pic3:RemnantDisks} we show the corresponding iso-density contours and overplot in
color all stellar particles (with individual masses of $m_\mathrm{star} = 
6.9 \times 10^4 M_{\odot}$) formed in the last $ \tau < 15 \Myr$ (blue), $15 \Myr < \tau <
  50 \Myr$ (green), and  $50 \Myr < \tau < 100 \Myr$ (red). In regions of currently high gas densities 
the very young stars (blue) form predominantly at the centers, in the overlap region, as well as in the 
spiral features around the disks similar to the observed system
\citep{WhitmoreEtAl1999AJ,WangEtAl2004ApJS}, save
the fact that the star formation in the centers seems to be much more
pronounced in our simulation (see below). However, the overlap region is almost 
devoid of stars older than $50 \Myr$ (red) indicating that the overlap starburst is a very recent 
phenomenon. Simulating the system further in time we find that the total duration
of the off-center starburst is no longer than $\approx 20 \Myr$.

\citet{2009ApJ...699.1982B} derived SFRs in the nuclei of NGC
  4038 ($0.63 \Msun \yr^{-1}$) and NGC 4039 ($0.33 \Msun \yr^{-1}$),
  and a total SFR of $5.4 \Msun \yr^{-1}$ for 5 infrared peaks in the
  overlap region. Comparing these values to simulated SFRs of $2.9
  \Msun \yr^{-1}$ (NGC 4038) and $2.8 \Msun \yr^{-1}$ (NGC 4039) in
  the galactic nuclei (defined as the central $\kpc$), together with
  $1.0 \Msun \yr^{-1}$ in the overlap region, we find that our
  simulation still falls short of producing the most intense starburst
  in the overlap compared to only modest star formation in the
  nuclei. We note, however, that we find a ratio
  $(\mathrm{SFR}_\mathrm{overlap}/\mathrm{SFR}_\mathrm{nuclei})$ of a
  factor of $\sim60$ (!) higher than reported in a previous Antennae
  model \citep{MihosBothunRichstone1993ApJ}.
The total SFR of $8.1 \Msun \yr^{-1}$ measured from the
SPH particles is in good agreement with the range of observed values
between $5 - 20 \Msun \yr^{-1}$
\citep[e.g.][]{ZhangFallWhitmore2001ApJ}. 

\begin{figure}
\centering 
\epsscale{1.}
\plotone{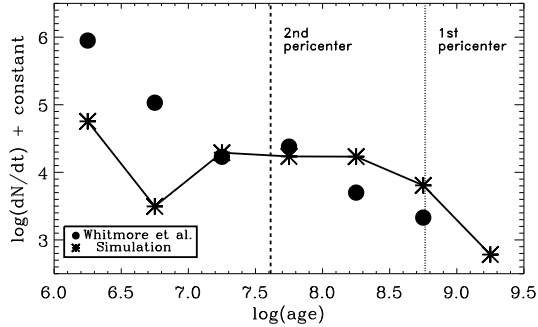}
\caption{Formation rate of stellar particles versus age for our simulation 
  (stars and solid line). 
  Vertical lines indicate the time of first (dotted) and second
  (dashed) pericenter. The observed cluster formation rate from
  \citet{WhitmoreChandarFall2007AJ} is given as filled circles.}
\label{pic4:SFH}
\end{figure}

In Fig.\ref{pic4:SFH} we
plot the formation rate of stellar particles within $18 \kpc$ against
their age (solid line). We find a significant increase of the SFR
after the first and in particular after the second  pericenter (dotted
and dashed horizontal lines). Assuming the simulated SFR to be
directly proportional to the cluster formation rate we compare the
simulated SFR to
observations of the age distribution of young star clusters
\citep{FallChandarWhitmore2005ApJ, WhitmoreChandarFall2007AJ}, using the same time binning 
of $0.5$ dex in $log(\tau\mathrm{[yr]})$. We find that our simulated data are in 
very good agreement exhibiting a similarly good  match to the observed
cluster formation rate as found by \citet{BastianEtAl2009ApJ...701..607B} who
compared to the \citet{MihosBothunRichstone1993ApJ} Antennae
model. This model predicted a nearly constant formation rate for ages
$\tau < 100 \Myr$. However, in contrast to the
\citet{MihosBothunRichstone1993ApJ} model, we find an additional significant
increase in the formation rate of young stellar populations at ages
$\tau \lesssim 10\Myr$ induced by the recent second encounter. Despite
the increase, the predicted formation rates of young clusters are still an order
of magnitude lower than observed. Further investigations have to show
whether this discrepancy originates from still uncertain details of
the star formation model or explicit effects of the early disruption and evolution of massive
clusters \citep[``infant mortality'', see 
e.g.][]{WhitmoreChandarFall2007AJ,BastianEtAl2009ApJ...701..607B}
which were not included in our model.

In this simulation, and other simulations in our parameter study with
similar central properties, we only find prominent star formation in the
overlap region for a very short period of time after the second encounter, 
lasting for only $\sim20 \Myr$. 
In addition, stellar feedback is required to prevent the rapid consumption 
of gas by star formation at earlier times. In a comparison 
run without stellar feedback most of the gas is consumed efficiently after the 
first pericenter and not enough gas is left over to form the overlap starburst after the 
second encounter. Thus, a central conclusion of our study is that the strong 
localized, off-center starbursts observed in the overlap region stems from a 
short-lived transient phase in the merging process associated with the recent second encounter. 


\section{Discussion} 
\label{discussion}

The new numerical model for the Antennae galaxies presented in this Letter 
improves on previous models in several key aspects. We find an excellent 
morphological and kinematical match to the observed large-scale morphology 
and \HI velocity fields \citep{HibbardEtAl2001AJ}. In addition, our model
produces a fair morphological and kinematical representation of the observed central region. 
A strong off-center starburst naturally develops in the 
simulation - in good qualitative and quantitative agreement with the observed extra-nuclear
star-forming sites \citep[e.g.][]{MirabelEtAl1998A&A,WangEtAl2004ApJS}. 
This is a direct consequence of our improved merger orbit. All previous studies using traditional orbits 
failed to reproduce the overlap starburst \citep [see e.g.][]
{KarlEtAl2008AN....329.1042K}. The exact timing after the second encounter shortly before the final merger ensures that the galaxies are close
enough for the efficient tidally-induced formation of the overlap region.
The formation of the extra-nuclear starburst is likely to be supported by compressive tidal forces which 
can dominate the overlap region in Antennae-like galaxy mergers during close encounters 
\citep{2008MNRAS.391L..98R,2009ApJ...706...67R}. 
Energetic feedback from supernovae prevents the depletion of gas by star formation at earlier merger stages and ensures 
that by the time of the second encounter enough gas is left over to
fuel the starburst. Simulating the system with an identical orbit,
but now employing an isothermal EQS without feedback from  supernovae ($q_\mathrm{EQS}=0$)
resulted in most of the gas being depleted by star formation at earlier phases of the 
merger, i.e. during the first encounter.

Our model predicts that the observed off-center starburst is a transient feature with a very short lifetime
($ \approx 20$ \Myr) compared to the full merger process ($\approx 650
\Myr$  from first encounter to final merger). This fact serves as a plausible explanation 
for why such features are rarely observed in interacting galaxies \citep{XuEtAl2000ApJ}. 
However, the observed puzzling gas concentration between the two
nuclei of the
\object{NGC 6240} merger system might be of a similar origin (\citealp[Engel in
prep. 2010]{1999ApJ...524..732T}) suggesting that the
Antennae overlap region, although rare, is not a unique feature.

In addition, our improved model can serve as a solid basis and testbed for further theoretical studies 
of the enigmatic interacting NGC 4038/39 system. For example, the overlap region in the Antennae 
is dominated by molecular gas, which we do not model in the simulation presented here. 
Given that we now have a dynamically viable method for forming the overlap
region, detailed investigations of the molecular gas formation process can be
undertaken using improved theoretical models \citep[e.g.][]{2008ApJ...680.1083R,2009ApJ...707..954P}.
In a first application using this new orbital configuration we have been able to qualitatively and 
quantitatively reproduce the magnetic field morphology of the Antennae
galaxies \citep{2009arXiv0911.3327K}. 

Finally, accurate modeling of nearby interacting systems also provide unique
insights into the merger dynamics and timing of observed merger systems. The
Antennae galaxies are traditionally in the first place in the classical Toomre
sequence which orders galaxies according to their apparent merger stage 
\citep{Toomre1977egsp.conf..401T} with the Mice (\object{NGC 4676})
being between their first and second pericenter \citep{Barnes2004MNRAS} and
thus in second place behind the Antennae. According to our proposed model the
Antennae galaxies are in a later merger phase, after the second pericenter.
As a consequence the Antennae would lose their first place, and thus requiring 
a revision of the classical Toomre sequence.

\begin{acknowledgements}
This work was supported by the DFG priority program 1177 
and the DFG Cluster of Excellence ``Origin and Structure of the
Universe''. We would like to thank J. Hibbard, N. Bastian, B. Whitmore
and M. Fall for valuable discussions on the manuscript. F. Renaud is a member
of the IK I033-N Cosmic Matter Circuit at the University of
Vienna.
\end{acknowledgements}

\end{document}